\documentclass[aip,pof,amsmath, amssymb,
12pt,floatfix]{revtex4-1}

\usepackage[utf8]{inputenc}
\usepackage{graphicx}
\usepackage{dcolumn}
\usepackage{bm}
\usepackage[mathlines]{lineno}
\usepackage{amsmath}

\usepackage{placeins}
\usepackage{longtable}

\usepackage{xr}
\makeatletter

\usepackage[caption=false]{subfig}

\newcommand{\phantomsubfloat}[1]{
    {
        \captionsetup[subfigure]{labelformat=empty}
        \subfloat[][]{#1}
    }%
}

\begin{document}

\title{Diffusion-driven deposition model suggests stiffer gels deposit more efficiently in microchannel flows}

\author{Barrett T. Smith}
\affiliation{Chemical Engineering, Northeastern University, Boston, MA 02115, USA}

\author{Sara M. Hashmi} 
\email[Corresponding Author: ]{s.hashmi@northeastern.edu}
\affiliation{Chemical Engineering, Northeastern University, Boston, MA 02115, USA}
\affiliation{Mechanical and Industrial Engineering, Northeastern University, Boston, MA 02115, USA}
\affiliation{Chemistry and Chemical Biology, Northeastern University, Boston, MA 02115, USA}

\begin{abstract}
The behavior of crosslinking polymer solutions as they transition from liquid-like to solid-like material in flow determines success or failure in several applications.  Dilute polymer solutions flow easily, while concentrated polymers or crosslinked polymer gels can clog pores, nozzles, or channels.  
We have recently described a third regime of flow dynamics in polymers that occurs when crosslinking happens during flow: persistent intermittency.  When a dilute alginate solution meets calcium at a Y-shaped microfluidic junction, a persistent and regular pattern of gel deposition and ablation emerges when driven at a constant volumetric flow rate. Chemical concentrations and flow rate control both the gel deposition and critical shear stress required to ablate the adhered gel. 
In this work, we provide an analytical framework to quantitatively describe the intermittent behavior as resulting from diffusively driven deposition in a high Peclet number flow. Fitting the experimental data shows that higher component concentrations lead to more efficient deposition and more swollen gels. Increasing the flow rate increases the deposition rate, but the resulting gels are much less swollen.
Ablation occurs when applied shear stresses overcome either the adhesive energy of the gel or its yield stress. The shear stress required at ablation decreases with increased component concentrations.
By correlating the results of the analytical analysis with bulk rheology measurements, we find that deposition efficiency increases with the stiffness of the gel formed in flow.  Softer gels withstand higher shear stresses before ablation. Both deposition efficiency and gel stiffness increase in flow conditions nearing complete clogging.

\end{abstract}

\maketitle

\section{Introduction}

The flow behavior of polymer solutions in microchannels dictates the success or failure in many systems. Dilute polymer solutions typically flow easily, while concentrated or cross-linked polymer solutions may clog. Bulk rheology describes the flow characteristics of polymer solutions. Properties such as shear thinning and yield stresses can describe how well a solution is likely to flow through microchannels, such as nozzles for three-dimensional (3D) printing \cite{Liu2023OnPrintability, Townsend2019FlowBioprinting}.  However, the flow of dilute polymer solutions with active cross-linking is not well described by bulk rheology.  Regions of localized gelation on the micro-scale may have different material properties from the bulk substance \cite{Geraud}.  Localized gelation also represents inhomogeneities that may be difficult to characterize.  The manner in which actively cross-linking solutions behave in micro-confined flow is relevant to many biological situations, such as blood clotting \cite{Fogelson} and biofilm formation \cite{Drescher}.  Elsewhere in the nature, in the context of petroleum engineering, properties of gelling polymer systems are optimized to block pores, control water content, and generally assist reservoir development and oil recovery \cite{liu2020investigation, tian2023gel, liu2022high}.  Manufacturing processes such as certain types of 3D printing \cite{Piras2020MulticomponentBioinks}, and microfluidic fabrication of hydrogels \cite{Hati, utech2015,martino2016controllable} rely on careful control of gelling systems to achieve appropriate material properties for a particular application.  These synthetic applications seek to avoid clogging, and typically also aim to avoid instantaneous cross-linking in small channel flows \cite{Hati}.  Despite the importance of polymer gelation in confined channels, more emphasis is typically placed on the material properties of the resulting gel.  The dynamics of gelation in flow, and its impact on the flow, remains underexplored.

Arguably the situation in which crosslinking polymer clogging dynamics is most well understood lies in the formation of blood clots \cite{Fogelson}. The final steps of the coagulation cascade include polymer crosslinking and deposition in confined flow. In blood coagulation, a complex series of regulatory proteins produces thrombin under appropriate conditions. Thrombin then converts the blood protein fibrinogen into insoluble fibrin. Fibrin molecules form fibers which aggregate and form a hydrogel structure, sometimes called a fibrin mat, that deposits on the channel wall. 
Many studies, both computational and experimental, have characterized how the flow and chemical environment influences the gel structure and strength of these fibrous mats.
High shear rates result in smaller fibrin fibers which align with flow \cite{gersh_flow_2010, onasoga2014thrombin}. High shear rates also restrict the protrusion of fibrin gels into the lumen of the vessel \cite{onasoga2014thrombin}. Since the elastic modulus of fibrin gels correlates with the fiber size, gels formed in high shear rate conditions are weaker than those formed at low shear rates \cite{ryan1999structural}.
Increased fibrin concentrations are associated with increased fiber density, shorter fiber length, smaller average hydrogel pore size, and thicker fibers \cite{garyfallogiannis_fracture_2023}. 
The effect of these changes is an increase in the resistance to fluid permeability within the gel.  This reduction in permeability in turn increases the toughness of the gel beyond the increase in toughness associated with the higher concentrations of solids alone.
Blood is a very complex organ with many important components beyond fibrin. Therefore, many studies investigate the interplay of the various factors and the influence of convective transport on more complicated structures and interactions even beyond those found in the fibrin mat.  In blood coagulation and the formation of clots inside vessels, the degree of complexity found in experimental and theoretical models is necessary to describe biological implications.

Controlling hydrogel formation in microfluidic flow is also important in research investigating the formation of hydrogel membranes in microfluidic devices \cite{ly2021flow}.
In these devices, hydrogels form where a polymer and a crosslinker mix downstream of a junction in stable flow. The gel grows to form a wall between the two streams. This wall can be used as a membrane between the two streams, or to trap cells and expose them to chemical gradients. 
Because of the potential biological applications, many of these devices use bio-derived hydrogels such as alginate and chitosan \cite{braschler2005gentle, bazargan2008formation, cheng2012biofabrication, ding2015chip, hu2020modulating, jia2020microfluidic, johann2007microfluidic, rosella2021microfluidic, correa2020microfluidic}.
Physical characterization of these membranes is limited by the difficulty of assessing membrane properties \textit{in situ} \cite{ly2021flow, ding2015chip}. Assessment relies primarily on microscopy to reveal gel formation, structure, and permeability, though additional tests can describe the gel adhesion to the channel wall and the volume swelling ratio of the hydrogel \cite{ding2015chip, hu2020modulating, luo2010situ, luo2014air, rosella2021microfluidic, sun2014situ}.
Controlling the area of reaction between polymer and crosslinker using channel geometry and careful control of flow is critical to avoiding undesirable deposition and clogging within these microfluidic devices \cite{braschler2005gentle, cheng2012biofabrication, ding2015chip, gargiuli2006microfluidic, jia2020microfluidic, johann2007microfluidic, luo2010situ, luo2014air}.
Varying component concentrations and flow rates have fairly consistent effects across materials and experimental setups. 
Increasing the concentration of either the polymer or crosslinker leads to larger and faster growing gel deposits \cite{braschler2005gentle, bazargan2008formation, jia2020microfluidic, correa2020microfluidic}.
Gels membranes tend to be thinner and grow more slowly when formed at faster flow rates \cite{bazargan2008formation, dabiri2023numerical, jia2020microfluidic, rosella2021microfluidic, johann2007microfluidic}, although some experiments show no consistent effect of changing flow rates on either membrane size or growth rate \cite{cheng2012biofabrication, luo2010situ, rosella2021microfluidic, ding2015chip, jia2020microfluidic}. Faster flow rates also contribute to high shear stresses which may remove gel material from the membrane or the membrane from the channel wall \cite{ding2015chip, gargiuli2006microfluidic}. Gels formed in faster flow rates  also have smaller volume swelling ratios \cite{rosella2021microfluidic}. 
Analytical descriptions of these techniques use component diffusion around and through the developing gel to explain gel growth \cite{braschler2005gentle,ding2015chip, gargiuli2006microfluidic, johann2007microfluidic, correa2020microfluidic}. 
However, these models describe gel growth in terms of changes in microscopy rather than externally measurable flow parameters. Further, they do not describe any relationship between gel properties and gel growth in these flows.

To more deeply study issues arising from polymer crosslinking in flow, we analyze one example of a minimal set of ingredients leading to rich dynamic phenomena: the biopolymer alginate mixes with its calcium crosslinker at a Y-shaped microfluidic junction \cite{smith2024}. At high concentrations of alginate and calcium, this channel clogs. At low concentrations, the channel exhibits uninhibited flow. In between these two regions, however, we observe an interesting and persistent intermittent deposition/ablation pattern. As the alginate and calcium mix, cross-linked alginate deposits on the channel wall. The deposited gel grows to partially occlude the channel, requiring an increase in driving pressure to maintain constant flow rates. At a certain critical pressure, the entirety of the deposited gel ablates from the wall and is flushed from the device. Gel begins depositing again, and the phenomenon repeats itself as long as flow continues. Both the extent of deposition and the frequency of ablation can be controlled by the concentrations of calcium and alginate and by the flow rate \cite{smith2024}. 
The persistent intermittent pattern provides an opportunity to study large numbers of deposition events. 
We analyze pressure trace data using an analytical diffusion driven deposition model and an analysis of shear stress to describe the efficiency of gel deposition and estimate gel properties as a function of chemical and physical conditions. 
We pair these \textit{in situ} observations and analytical descriptions with bulk rheological measurements. 
Our results demonstrate that stiffer gels deposit more efficiently but ablate at lower shear stresses than softer gels.

\section{Materials and Methods}

\subsection{Materials}

Sodium alginate and sodium chloride are purchased from Sigma-Aldrich. Calcium Chloride is purchased from VWR. The PDMS is purchased as the Sylgard 184 Silicone Elastomer Kit from DOW Silicones Corporation. All chemicals are used as received. Stock solutions of sodium alginate, calcium chloride, and sodium chloride are prepared in DI water (Milli-Q Advantage A10) above their target concentrations. 
Appropriate amounts of sodium chloride stock solution are added to the alginate and calcium solutions to achieve ratios of 1:25 (mg/mL : mM) and 100:2.5 (mM : mM) respectively before each dilution to the target concentrations. 
Alginate concentrations vary from 0.02 to 0.2 mg/mL, corresponding to sodium chloride concentrations of 0.5 to 5mM respectively.
Calcium concentrations vary from 10 to 200mM.

For microscopic imaging, alginate is covalently tagged with a fluorescent label \cite{scheja2017glucose}. To tag the alginate, a quantity of 255.2mg sodium alginate, 18.5mg fluoresceinamine (Thermo Scientific), and 35.9mg N-(3-Dimethylaminopropyl)-N-ethylcarbodiimide hydrochloride (Sigma) are dissolved in 15mL phosphate buffered saline. The reaction is stirred for 24 hours at 25$^\circ$C. The sample is then dialyzed in de-ionized water for one week with the bath solution replaced daily. Lastly, the dialyzed sample is lyophilized and kept at -20$^\circ$C until use.

\subsection{Microfluidic Device}

The device used in these experiments consists of a microfluidic Y-intersection (Fig. ~\ref{fig:device}). Inlet channels, 40 $\mu$m wide and 6 mm long, meet at a 60$^{\circ}$ angle. The outlet channel is also 40 $\mu$m wide, but is longer at $L=10$ mm. All channels are 25 $\mu$m high. Soft lithography is used to manufacture these devices according to standard methods \cite{McDonald}.
However, rather than core an outlet channel for outlet tubing, the last millimeter of the outlet channel is cut, which allows the effluent to discharge directly onto the slide. This is done to prevent the build-up of alginate gel in the outlet well.  

\subsection{Flow Tests}  

A pressure controller (Fluigent Flow EZ) with an attached flow rate meter is used to set constant flow rates for experiments. 
The flow rate ratio of alginate and calcium channels is held constant at 4:1.
For experiments varying calcium $C_{Ca^{2+}}$, the individual flow rates $Q$ are set at 0.96 $\mu$L/min and 0.24 $\mu$L/min for alginate and calcium streams respectively.
For experiments varying the gel concentration $C_{Gel}$, the individual flow rates are set to 4.8 $\mu$L/min and 1.2 $\mu$L/min.
For experiments varying flow rates, $Q$ varies from 0.96 to 9.6 $\mu$L/min for the alginate stream, and 0.24 to 2.4  $\mu$L/min for the calcium stream. 
We refer to flow rates by the total sum of the individual flow rates, $Q_T$.
Despite the variation in driving pressure due to the accumulation of gel, $Q$ remains constant \cite{smith2024}.
The tubing is flushed with the desired solution before the being connected to the microfluidic device. Pressure data is recorded every 100ms for the duration of the experiment, which is typically at least 15 minutes. 

All flow tests are performed at high Peclet number, where $\text{Pe} = Q/D R_0$ indicates the balance between convection and alginate diffusion, where $R_0$ is the radius of the channels before any deposition occurs.  $D$ is the Brownian diffusion constant $D=k_BT/6\pi \mu a$ where $\mu$ is the viscosity of the suspending fluid and $a$ the effective radius of the alginate molecules before crosslinking, which we approximate as $\sim4$ nm.  For all flow tests, the range of $\text{Pe}$ for alginate is $\sim 2\times 10^4$ to $2\times 10^5$.  It is worth noting that these values likely underestimate $\text{Pe}$: as alginate crosslinks, the effective size of the diffusing moiety may increase significantly, causing $D$ to decrease.  Reynolds numbers are on the order of $\text{Re}\sim O(1)$.

\subsection{Fluorescent Microscopy}

We record gel deposition using an inverted microscope (Leica DMi8) at 63× magnification.  The fluorescein labeled alginate has a peak excitation wavelength of 495 nm and a peak emission wavelength near 520 nm. A video of gel deposition is available online. 
In the video, the narrow focal plane of the microscope is set a few microns above the glass slide to enable visualization of gel as it grows into focus. 
The video is collected during a flow test with $C_{Alg} = 0.1$mg/mL and $C_{Ca^{2+}}=100$mM through a channel with a cross-section of $100\mu \text{m} \times 50 \mu \text{m}$,  at $Q_T = 2\mu\text{L/min}$, with a flow rate ratio of 1:1. The video is slowed down by a factor of 4 from real time.

\subsection{Rheology}

The rheology of alginate solutions and gels is measured using a Discovery HR-3 rheometer from TA Instruments.  For solutions used in the flow tests, shear rate sweeps reveal the expected Newtonian behavior, and the viscosity of the alginate solutions is less than 1.1 cP for every concentration used in the microfluidic experiments.  As such, differences in viscosity is considered negligible in the analysis that follows.

To characterize the crosslinked gel that forms in flow, rheology of bulk alginate solutions is measured. Samples are prepared in bulk: sodium alginate is dissolved at 1 and 2 mg/mL  with solutions of 25 mM NaCl and 50 mM NaCl, respectively.  Calcium chloride solutions of different concentrations are added dropwise to the alginate solutions while vortexing to ensure mixing.  The ratio of alginate solution to calcium solution is maintained at the 4:1 volume ratio that occurs in the microfluidic device. To perform bulk rheology, we increase the alginate concentration by an order of magnitude compared to the concentration used in the flow tests. The ratio of calcium to alginate investigated in bulk rheology overlaps the ratios used in the flow tests.  In flow tests, for $C_{Alg}=0.1$ mg/mL, calcium is added at 10, 20, 100, and 200 mM.  For bulk rheology of alginate gels, calcium is added at 10, 20, 35, 50, 100, and 200 mM.  

Prepared solutions are added to a concentric cylinder geometry for rheology measurements for small amplitude oscillatory shear (SAOS) testing, consisting of an amplitude sweep followed by a frequency sweep.
Prior to each step, the sample is conditioned at a shear rate $\dot{\gamma}=50$ 1/s for 1 minute. 
The amplitude sweep is measured at an angular frequency $\omega=5$ rad/s.  All samples exhibit a linear regime with constant modulus at strains below 0.1\%.
The frequency sweep is measured at an oscillatory strain of 0.1\%, which is within the linear regime found from the amplitude sweep for all samples.

\section{Flow Test Results}


\begin{figure}
    \includegraphics[width=\columnwidth]{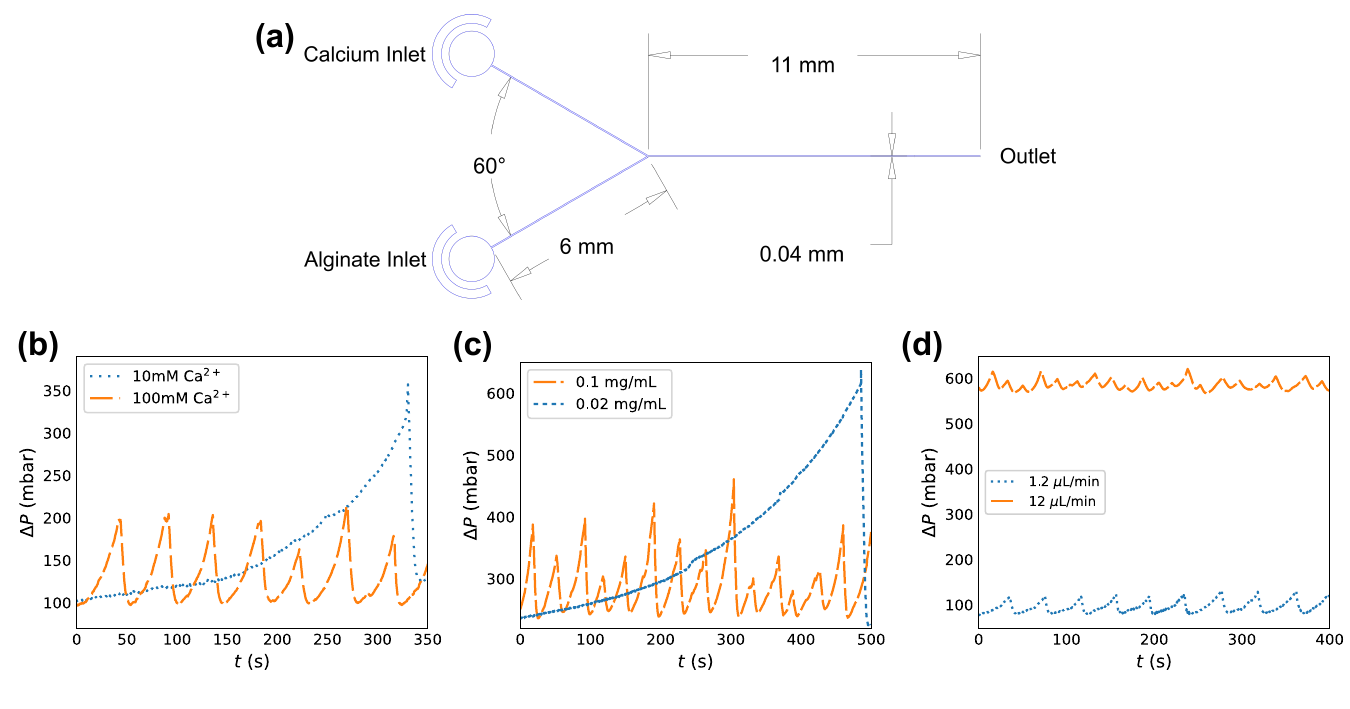}

    \phantomsubfloat{\label{fig:device}}
    \phantomsubfloat{\label{fig:Calcium Raw}}
    \phantomsubfloat{\label{fig:Concentrations raw data}}
    \phantomsubfloat{\label{fig:Flow rate raw data}}
    \vspace{-2\baselineskip}

    \caption[Summary of Experimental Data]{
    (\subref{fig:device}) Device schematic: two inlet channels, 6mm in length, meet in a Y-shaped junction at an angle of 60$^\circ$.  All channels are 40$\mu$m wide and $\sim30 \mu$m tall. 
    (\subref{fig:Calcium Raw}), (\subref{fig:Concentrations raw data}), (\subref{fig:Flow rate raw data}) Representative raw pressure traces from calcium experiments, total concentration experiments, and flow rate experiments respectively.
    }
    
    \label{fig:Set up and Initial Data}
\end{figure}

When alginate and calcium mix at constant flow rates in a microfluidic Y-junction, gel deposits along the fluidic channel wall.  As long as the concentrations of alginate and calcium are not sufficiently high to completely clog the device, most flow tests results in an interesting and persistent intermittent pattern of gel deposition followed by gel ablation \cite{smith2024}. Fig.~\ref{fig:Set up and Initial Data} shows the device geometry as well as several pressure traces of this intermittent flow obtained at different conditions.  Each of these pressure traces demonstrates periodic behavior where, starting at $t_{i,0}$ and $\Delta P_0$, alginate gel deposits on the channel wall. As the alginate deposit grows, it obstructs flow, requiring an increase in driving pressure to maintain a constant flow rate. At a certain level of occlusion, at time $t_{i, Abl}$ and pressure $\Delta P_{Abl}$, shear stress ablates the gel from the wall and the gel is flushed from the device. The driving pressure in the channel returns to baseline.  The gel then begins to deposit again, and the pattern repeats as long as flow continues. 

The driving pressure at ablation, $\Delta P_{Abl}$, and the frequency of ablation, $f_{Abl}$ are both controlled by chemical and physical parameters. The variation in behavior associated with changing the concentration of cross-linker $C_{Ca^{2+}}$, the concentration of gel $C_{Gel}$, and the flow rate $Q_{T}$ are illustrated in figures \ref{fig:Calcium Raw}, \ref{fig:Concentrations raw data}, and \ref{fig:Flow rate raw data} respectively. $C_{Gel}$ corresponds to the concentration of alginate at a fixed crosslink ratio, 100 mM calcium per 0.1 mg/mL sodium alginate.  Calcium cross-links alginate. Therefore, increasing the concentration of calcium $C_{Ca^{2+}}$ increases the degree of crosslinking. Figure \ref{fig:Calcium Raw} shows that decreasing $C_{Ca^{2+}}$ results in larger peaks, $\Delta P_{Abl}$, and a much lower frequency of ablation, $f_{Abl}$. Increasing the concentration of gel, while keeping the ratio of alginate to calcium constant, shows a similar trend to that of crosslink density.  Figure \ref{fig:Concentrations raw data} shows that as the gel concentration decreases, ablation events decrease in frequency while increasing in the magnitude of driving pressure at ablation. Increasing the flow rate increases the baseline driving pressure ($\Delta P_0$), but the absolute peak size and frequency of ablation are similar despite an order of magnitude difference in $Q_T$, as seen in \ref{fig:Flow rate raw data}. While the absolute difference in driving pressure between baseline and clog ablation ($\Delta P_{Abl} - \Delta P_0$) and the frequency of ablation $f_{Abl}$ do not vary significantly with $Q_T$, it is worth considering that $Q_T$ controls both the baseline driving pressure and the baseline wall shear rate. Thus, the normalized pressure at ablation $\Delta P_{Abl}/\Delta P_0$ decreases as $Q_T$ increases.  The normalized peak size is $\Delta P_{Abl}/\Delta P_0\sim$ 162\% at at 1.2 $\mu$L/min, and decreases to 109\% at at 12 $\mu$L/min.

\section{Microscopy of gel growth and structure}
\begin{figure}
    \centering
    \phantomsubfloat{\label{fig:Fractal Pic}}
    \phantomsubfloat{\label{fig:Capillary Pic}}
    \phantomsubfloat{\label{fig:Confocal Pic}}
    \vspace{-2\baselineskip}
    
    \includegraphics[width=0.5\columnwidth]{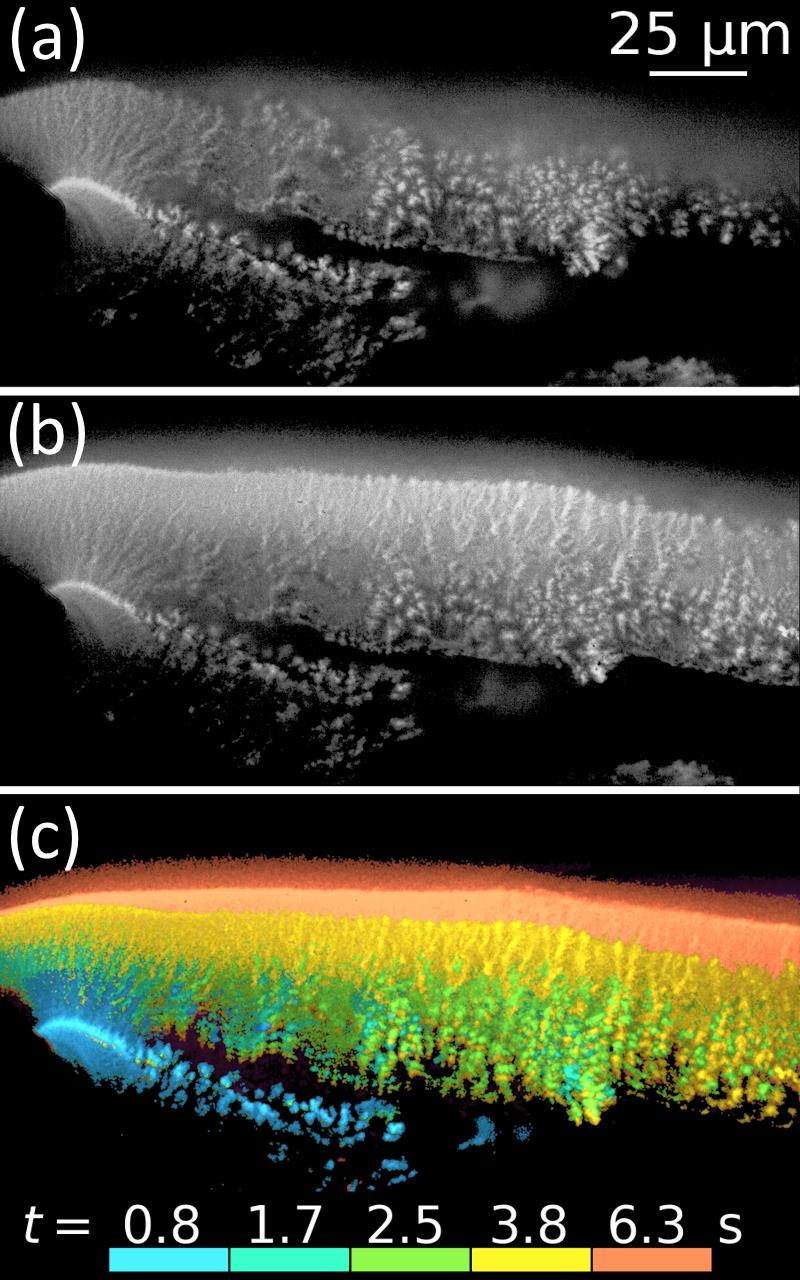}
    
    \caption[Microscopy.]
    {(\subref{fig:Fractal Pic}) Fractal growth seen on the gelation front.
    (\subref{fig:Capillary Pic}) Striations parallel with the direction of gel growth suggest capillary formation. (c) A color-coded overlay of individual images shows the growth of the deposit over time.
    Video taken in a $100\times50\mu\text{m}$ channel, $C_{Alg} = 0.1 \text{mg/mL}, C_{Ca^{2+}}=100\text{mM}$. Multimedia available online. }
    \label{fig:Microscopy}
\end{figure}

Fig.~\ref{fig:Microscopy} (Multimedia available online) shows microscopy of the gel during gel deposition. The still images in Fig.\ref{fig:Microscopy} are taken from the video at timestamps $\sim 4$s and $\sim 7$s, respectively. The video is slowed by a factor of four compared to real time. Solutions flow from left to right: the alginate stream ($C_{Alg} = 0.1$mg/mL) enters from the upper left and the calcium stream ($C_{Ca^{2+}}=100$mM) enters from the lower left. The alginate is tagged with a fluorescent label and appears bright in the video.  The video starts with a clean channel, although a small portion gel from a previous deposition/ablation event remains at the junction and near the side wall. This remaining gel does not obstruct the flow significantly: the driving pressure begins at baseline.  In the beginning of the video, alginate gel grows from the glass slide below the focal plane. This growth starts to appear as unfocused light. As more alginate deposits, the gel comes into focus, and interesting fractal patterns emerge, as shown in Fig.\ref{fig:Fractal Pic}. As deposition continues, gel from above the focal plane contributes to the out of focus signal and observation of the detailed gel structure becomes difficult. In addition to growing up from the glass slide, alginate gel grows towards the side walls. This can be seen clearly in the time-overlay shown in Fig.\ref{fig:Confocal Pic}.  While not shown in the video, similar gel growth also occurs on the ceiling of the channel, so that the gel grows from the top and the bottom of the channel.  Eventually, the gel growing inward from top and bottom merge in the middle of the channel to form one continuous gel.  In this growth, a striated structure grows parallel with gel growth and perpendicular to the direction of flow, as seen in Fig. \ref{fig:Capillary Pic}. This structure may be capillary fingers similar to those seen in alginate gel formed in quiescent conditions \cite{treml2003theory}.

Taken together, the images and video show gel growth consistent with our interpretation of the pressure trace data. Gel deposits on the walls of the device and grows inward to obstruct flow. The obstruction is significant and reduces the effective cross-sectional area of the channel. Thus a higher driving pressure is required to maintain a constant flow rate. To maintain constant volume flow, the linear velocity increases as the effective cross-section decreases until the increased shear stress is high enough to remove the gel the walls of the device.

\section{Analytical Models}

To describe gel deposition, growth and ablation phenomenon, we employ multiple analyses of the pressure trace measurements. We first use a convection-diffusion transport model to describe the deposition behavior.  We then analyze the point of ablation to investigate the amount of stress required to pull the gel clogs off the channel walls.  Together, these analyses inform us about both the growth of the gel and its properties inside the channel. The discussion of ablative stress and gel properties then leads into the discussion of the bulk rheology of the gels.

The analyses below assume pipe flow, whereas the fluidic devices have a cross-section that are nearly square.  We use the Poiseuille equation to measure the effective radius of the devices.  In a pipe of radius $R$, or a channel with effective radius $R$, the pressure drop $\Delta P$ is proportional to $1/R^4$ by:
\begin{equation}\label{eq: Poiseuille}
    \frac{\Delta P}{L} = \frac{8\mu Q}{\pi R^4}
\end{equation}
where $L$ is the length of the pipe.  By flowing pure water through the devices over a range of pressure drops and measuring the resulting flow rate, we find that the effective radius of the $40\times25\mu$m channel is $R\simeq 19 \mu$m.


\subsection{Deposition Model}
\label{section:deposition model}

\begin{figure}
    \centering
    \phantomsubfloat{\label{fig:Raw Data for Example}}
    \phantomsubfloat{\label{fig:Example Data Collapse}}
    \phantomsubfloat{\label{fig:Iterative Model}}
    \phantomsubfloat{\label{fig:Data Transform}}
    \vspace{-2\baselineskip}
    
    \includegraphics[width=\columnwidth]{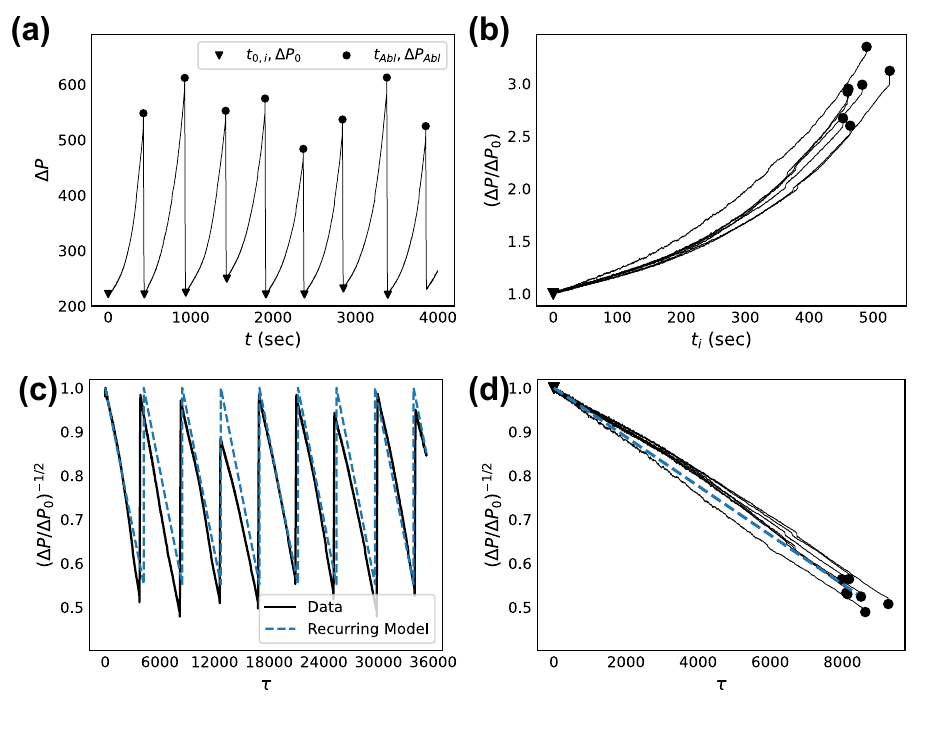}
    
    \caption[Initial Data - normalized and transformed.]
    { (\subref{fig:Raw Data for Example})  A representative pressure trace taken at $C_{Alg}=$ 0.02 mg/mL, and $C_{Ca2+}=100$ mM, which has a representative peak shown in Fig.{\protect\ref{fig:Concentrations raw data}}. 
        (\subref{fig:Example Data Collapse}) Each clogging event from (\subref{fig:Raw Data for Example}), isolated from its trough to peak, normalized by $\Delta P_0$ and shifted to $t_{0,i}=0$. 
        (\ref{fig:Iterative Model}) Data from (\subref{fig:Raw Data for Example}), transformed according to the deposition model Eq. \ref{eq: model}, plotted with the iterative model pressure trace from Eq. \ref{eq: itermodel}, where $\tau = Q (t - t_{0i}) / V$.
        (\subref{fig:Data Transform}) The same traces transformed and shifted to $\tau_{0,i}=0$.  
    }
    \label{fig:Initial Transform}
\end{figure}

To better understand the deposition phenomenon, we apply a diffusive flux deposition growth model derived from first principles for high-$\text{Pe}$ flows \cite{Hashmi}. Given flow rates on the order of $Q\sim1 \mu$L/min, all measurements are obtained at $\text{Pe}\sim 10^4$ or higher. In brief, the model assumes that the crosslinked alginate in a region within the diffusive boundary layer $\delta$ near the wall accounts for the deposit.  The diffusive flux $F$ of the depositing material toward the wall builds up there, growing inward and causing a decreases in radius. This mirrors the deposition behavior seen in microscopy.  The change in the cross-sectional area over time is given by:
\begin{equation}\label{eq: depo differential}
    \pi \frac{d(R_0^2-R^2)}{dt}= 2\pi R F
\end{equation}
\noindent where $R_0$ is the effective radius of the channel in the absence of deposit.  The flux $F$, having the same dimensions as a velocity, is determined by Fick's law: $F=-D \frac{d\phi}{dr}$, where $\phi$ represents the volume fraction of cross-linked alginate. We use cross-linked alginate because alginate itself does not deposit on the channel walls \cite{smith2024}. 
The scaling for $F$ is thus $D\phi/\delta$, where $\delta \sim R \text{Pe}^{-1/3}$ when $\text{Pe}$ is large and convection dominates over diffusion \cite{Acrivos}. Because $F\sim D\phi/\delta$ represents a scaling relation and not an equation, we write $F = kD\phi/\delta$.  The constant $k$ represents the efficiency of deposition, or the fraction of the total flowing material $\phi$ that has a flux toward the wall within the boundary layer $\delta$.  

By integrating equation \ref{eq: depo differential} and using the Poiseuille equation (\ref{eq: Poiseuille}) to relate the decrease in radius to an increase in pressure, we arrive at the following governing equation:
\begin{equation}\label{eq: model}
    \bar{R}^2= \left(\frac{\Delta P}{\Delta P_0}\right)^{-\frac{1}{2}} = 1 - B{\text{Pe}}^{-\frac{2}{3}}k\phi \tau
\end{equation}
where the radius is non-dimensionalized,
$\bar{R}=R/R_0$, and 
$B=\frac{2\pi L}{R}$
represents a geometric factor that is constant for all experiments within the same device geometry, and is 3307 for a $40\times25\mu\text{m}$ channel with a length of 10mm.  Time is rescaled to $\tau = Q(t-t_{i,0})/V$ where $t_{0i}$ is the trough between peaks in the pressure trace and $V=\pi R^2L$ the volume of the channel downstream of the junction. Dimensionless time $\tau$ has the meaning of pore volumes, or the number of times the channel has been completely filled with fluid. Because $B$ and $\text{Pe}$ are known, $k\phi$ represents a single fitting parameter.  While the concentration of alginate flowing through the channel is a known quantity, the fraction of the flowing alginate which crosslinks, $\phi$, is unknown, as is the amount of crosslinked alginate which deposits on the channel, $k\phi$. By incorporating the constant flow rate, $k\phi Q$ can also be used as a fitting parameter, interpreted as the volume flow rate of alginate deposited in the channel.

Equation \ref{eq: model} represents the increase in pressure during the duration of a single deposition event.  We can iterate this model by defining $\tau_{Abl}$ as the pore volumes required to reach ablation pressure at specific conditions. $\Delta P (\tau)$ can then be expressed simply:
\begin{equation}\label{eq: itermodel}
(\Delta P / \Delta P_0)^{-1/2} = 1- B \text{Pe}^{-2/3}k\phi\tau_{Abl}(\frac{\tau}{\tau_{Abl}}- i )
\end{equation}
\noindent using the floor function $i=\lfloor \frac{\tau}{\tau_{Abl}} \rfloor$ to define the integer index of each deposition-ablation event. In this way only the measured average ablation time is needed to describe $\Delta P$ for the duration of an entire flow test.

The model relies on several assumptions. We approximate the channel, with nearly square cross-section, as a pipe of circular cross-section, to develop a first-order approximation of the deposition behavior.  The scaling of the boundary layer $\delta$ and thus the flux toward the wall $F$ require high $\text{Pe}$, which is reasonable given the minimum $\text{Pe}\sim10^4$ across all experiments performed.  However, the model does not include chemical reaction kinetics: crosslinking depends on the interaction of calcium with alginate.  In bulk gels, calcium ion transport occurs by diffusion or active mixing.  In flow, convection also drives motion of calcium toward alginate.  Once in proximity, however, calcium crosslinks alginate instantaneously, even in microfluidic flows \cite{Hati, treml2000coupling, donati2023alginate}.  This suggests that crosslinking is limited by transport, but the reaction timescale itself is fast enough to be neglected here.  A third assumption of the model, which is also the most likely to break down, is that deposition happens uniformly both around the perimeter of the channel and along the length of the deposit. At the same time, the length of the deposit does not appear in the final formulation of the model in Eq. \ref{eq: model}.  That is, while $L$ appears both within the geometric parameter $B$ and the denominator of the rescaled time $\tau$, these two instances of $L$ cancel each other.

The implementation of this model leads to a testable hypothesis. 
If the assumptions of the model are satisfied, then the transformation suggested by Eq.~\ref{eq: model} should result in a linear relationship between $\left(\Delta P/\Delta P_0\right)^{-1/2}$ and dimensionless time $\tau$. Fig.~\ref{fig:Raw Data for Example} shows a pressure trace that exhibits eight individual deposition/ablation events over 4000 s of a flow test.  The troughs of each event are labeled with an inverted triangle, and the peaks with a circle.  Each event is isolated, shifted to $t_{0,i}=0$s, normalized by $\Delta P_0$, and then plotted together in Fig.~\ref{fig:Example Data Collapse}.  Next, time is non-dimensionalized by converting to $\tau = Q (t - t_{0i}) / V$, and pressure rescaled to $(\Delta P/\Delta P_0)^{-1/2}$ to match the form shown in Eq.~\ref{eq: model}. All eight traces from Fig.~\ref{fig:Raw Data for Example} are rescaled in this way and plotted in Fig.~\ref{fig:Data Transform}.  The troughs are still indicated by an inverted triangle, now located at (0,1), and the ablation points by a circle.  Indeed, each scaled trace in Fig.~\ref{fig:Data Transform} is a line with an intercept at $(\Delta P/\Delta P_0)^{-1/2}=1$ and a negative slope as a function of $\tau$.  The average dimensionless slope of the traces is $-1.01\times10^{-4}\pm 5.3\times10^{-6}$.  The blue dashed line has the averaged slope of the traces. The traces are largely linear, as indicated by  $R^2= 0.996\pm0.002$. Given the experimental conditions, namely $C_{Alg}=$ 0.02 mg/mL, $C_{Ca2+}=100$ mM, and $Q_T=1.2 \mu$L/min, measuring the slope suggests that $k \phi Q=0.12$ for this condition. Interestingly, gel deposits grow fairly slowly in $\tau$-space, reaching the point of ablation after hundreds or thousands of pore volumes.  In the example shown in Fig. 2d, $\tau$ reaches nearly 10,000 before ablation occurs. 

The agreement that the raw data exhibit when rescaled to the assumptions of the model suggests that diffusively driven deposition in high $\text{Pe}$ flows provides a very reasonable description of \textit{in situ} crosslinked alginate depositing on the walls of a microfluidic channel.  Interestingly, the model works remarkably well despite its geometric assumptions.  That is, the model assumes uniform deposition around the perimeter of the channel.  However, the microscopy images and video suggest that deposition occurs on the top and bottom of the device at the interface between the two inlet streams (Fig. \ref{fig:Microscopy}), but not on the sidewalls of the channel. Also, deposition at the junction occurs earlier than downstream deposition \cite{smith2024}. Nevertheless, we find linear behavior in the rescaled $(\Delta P / \Delta P_0)^{-1/2}$ traces across a variety of chemical and physical conditions, which we explore next. Supplemental Material Fig. S1 provides a histogram of all $R^2$ values for all rescaled pressure traces. The median $R^2$ value is 0.973 for 624 ablation events; 93.8\% of the events have an $R^2 > 0.9$.  Supplemental Material Fig. S1 also displays a few non-linear rescaled pressure traces and we explain their behavior. Further, we test the validity of the model in channels with different geometrical cross-sections.  Supplemental Material Fig. S2 shows that the pressure rescaling suggested by the model produces linear results in channels of different width:height aspect ratios, varying from 2:5 to 10:5.


\begin{figure}
    \centering
    
    \includegraphics{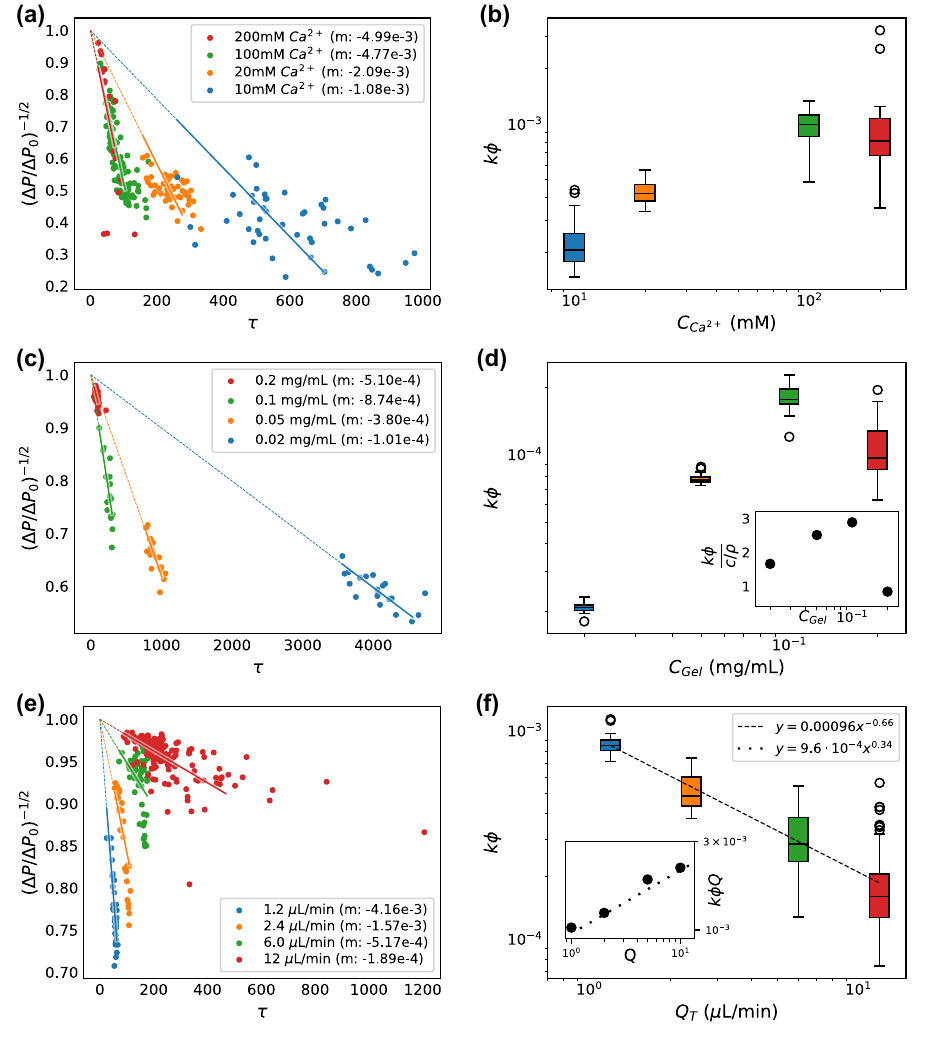}
    
    \phantomsubfloat{\label{fig:Calcium summary}}
    \phantomsubfloat{\label{fig:Calcium Slopes}}
    \phantomsubfloat{\label{fig:Concentrations summary}}
    \phantomsubfloat{\label{fig:Concentrations Slopes}}
    \phantomsubfloat{\label{fig:flow rate summary}}
    \phantomsubfloat{\label{fig:Flow rate Slopes}}
    \vspace{-2\baselineskip}
    
    \caption[Deposition Model]{
    (\subref{fig:Calcium summary}), (\subref{fig:Concentrations summary}), (\subref{fig:flow rate summary})
    Driving pressure at ablation transformed according to the deposition model for  calcium concentrations, total gel concentrations, and variable flow rates respectively.
    (\subref{fig:Calcium Slopes}), (\subref{fig:Concentrations Slopes}), (\subref{fig:Flow rate Slopes}) $k\phi$ values extracted from slopes of the best fits on the left.
    }
    \label{fig:Deposition Model}

\end{figure}

\begin{table}
    \caption{Comparing rates of deposition and swelling ratios at different experimental conditions.}
    \begin{center}
        \begin{tabular}{|| c   |c  |c  | c ||} 
            \hline\hline
            
            \textbf{$C_{Ca^{2+}}$ ($mM$)} &  Number of Events&$\mathbf{k\phi Q}$ ($nL/min$) &$S^*$\\
            \hline
            10 &  45&$0.27\pm0.076$&$17.23\pm4.92$\\ 
            \hline
            20  &  53&$0.52\pm0.074$&$33.42\pm4.78$\\
            \hline
            100 &  118&$1.18\pm0.216$&$76.5\pm13.95$\\
            \hline
            200  &  18&$1.26\pm0.858$&$81.20\pm55.42$\\
            \hline\hline

            \textbf{$C_{Gel}$ ($mg/mL$)}&  Number of Events&$\mathbf{k\phi Q}$ ($nL/min$) &$S^*$\\
            \hline
            0.02 &  20& $0.12\pm 0.007$ &$8.05\pm0.42$\\ 
            \hline
            0.05 &  19&$0.47\pm 0.027$ &$12.17\pm0.70$\\
            \hline
            0.1 &  17&$1.08\pm 0.162$ &$13.98\pm2.10$\\
            \hline
            0.2 &  54&$0.63\pm0.176$ &$4.08\pm1.13$\\
            \hline\hline
            
            \textbf{$Q_{T}$ ($\mu L/min$)} &  Number of Events&$\mathbf{k\phi Q}$ ($nL/min$) &$S^*$\\
            \hline
            1.2&  38&$1.03\pm0.109$  &$66.60\pm7.02$\\ 
            \hline
            2.4&  34&$1.24\pm0.245$  &$25.18\pm4.99$\\
            \hline
            6.0&  59&$1.87\pm0.656$  &$8.28\pm2.90$\\
            \hline
            12.0&  149&$2.17\pm0.915$  &$3.02\pm1.27$\\ 
            \hline\hline
        \end{tabular}
    \end{center}
    \label{table:flow rates kphiQ}
\end{table}

As observed in the pressure traces shown in Fig. 1, both the change in pressure gradient and the time to ablation vary significantly according to reaction and flow conditions \cite{smith2024}.  Rescaling every raw data trace in the manner described above allows us to explore the dependence of $k\phi$ on physical and chemical conditions.  We investigate 12 different flow conditions.  Figs.~\ref{fig:Calcium summary},~\ref{fig:Concentrations summary}, and ~\ref{fig:flow rate summary} show the results of re-scaling the pressure traces for each experimental condition, organized in sets to show the effect of varying calcium, varying gel concentration, and varying flow rates respectively. For \ref{fig:Calcium summary}, $C_{Alg}=$ 0.1 mg/mL and $Q_T=$ 1.2 $\mu$L/min.  For \ref{fig:Concentrations summary}, $C_{Ca^{2+}}=$ 100mM and $Q_T=$ 6 $\mu$L/min. For \ref{fig:flow rate summary}, $C_{Alg}=$ 0.1 mg/mL  and $C_{Ca^{2+}}=$ 100mM.  The pressure traces in each condition exhibit many individual deposition/ablation events totaling 624 (Table \ref{table:flow rates kphiQ}).  The rescaled pressure trace of each event is roughly a straight line from the point $\tau =0$, $(\Delta P / \Delta P_0)^{-0.5}=1$ to the plotted point.

For ease of viewing the rescaled results, instead of plotting in the same manner as in Fig.~\ref{fig:Data Transform}, we plot only the final point of each deposition/ablation event. Each point in Figs.~\ref{fig:Calcium summary},~\ref{fig:Concentrations summary}, and ~\ref{fig:flow rate summary} represents one deposition/ablation event at the point of ablation.  Points lower on the graph have a larger difference between baseline pressure and pressure at ablation, which corresponds to a larger obstruction of the channel. Points at higher $\tau$ values show longer periods of deposition before ablation. The average slope of all pressure traces at each condition is represented as a line superimposed on the data points. The slope of each line, as shown in Eq.~\ref{eq: model}, is proportional to the dimensionless deposition rate $k\phi$. The box-and-whiskers plots of Fig.~\ref{fig:Calcium Slopes},~\ref{fig:Concentrations Slopes}, and ~\ref{fig:Flow rate Slopes} show the $k\phi$ values at each of the corresponding conditions. In the box and whisker plots, the line inside the box indicates the median, while the box represents data from the 1st to 3rd quartile. The whiskers extend 1.5 times the interquartile range. All outliers beyond these limits are shown as individual data points. Absolute deposition rates can be determined by multiplying $k\phi$ by flow rate $Q$. These values, in units of nL/min, are found in Table~\ref{table:flow rates kphiQ}.

Fig.~\ref{fig:Calcium summary} and \ref{fig:Calcium Slopes} show that alginate is deposited more efficiently at higher concentrations of calcium. (Fig.~\ref{fig:Calcium Slopes}) shows that, at $C_{Ca^{2+}}=$ 10mM, $k\phi= 2.2 \pm 0.6 \times10^{-4}$. There is a 5 fold increase at $C_{Ca^{2+}}=$ 100mM;  $k\phi=9.8 \pm 1.8 \times 10^{-4}$.  However, this increase in deposition efficiency with calcium apparently stops there.  The rate of deposition is nearly the same between $C_{Ca^{2+}}=$100mM and 200mM. Interestingly, inspection of Fig.~\ref{fig:Calcium summary} shows that the rescaled pressure traces not only have similar slopes, but also persist for approximately the same amount of time, $\tau$, and even have similar values of $\Delta P$ at the point of ablation.

Figs.~\ref{fig:Concentrations summary} and \ref{fig:Concentrations Slopes}  show the effect of changing $C_{Gel}$, which is represented as $\phi$ in the model.  In Eq.~\ref{eq: model}, $\phi$ appears in the slope of the rescaled pressure as a function of $\tau$. Expectedly, therefore, \ref{fig:Concentrations summary}  shows the slope of the transformed pressure traces decrease in magnitude as $\phi$ decreases. Increasing gel concentration 5-fold from $C_{Gel}=0.02$ mg/mL, in blue, to $C_{Gel}=$ 0.1mg/mL, in green, results in a nearly 9-fold increase in $k\phi$, from $2.1 \pm 0.1 \times10^{-5}$ to $1.8 \pm 0.3 \times10^{-4} $. At $C_{Gel}=0.2$mg/mL however, in red, $k\phi$ drops to $1.1 \pm 0.3 \times10^{-4}$.  Given the overlap in the data sets at $C_{Gel}=0.1$mg/mL and $C_{Gel}=0.2$mg/mL, this behavior appears as though it may be similar to the plateau seen at $C_{Ca^{2+}}=$100mM in Fig. 3b.  Another possible explanation lies in the flow controller response time: when $C_{Gel}=0.2$mg/mL, ablation occurs frequently, with the time between ablation events ($\sim9$s) approaching nearly the response time of the flow controller ($<5$s). As such, measurements of $\Delta P_0$ at ablation may be artificially high.  This would cause the slope of $(\Delta P / \Delta P_0)^{-1/2}$ as a function of $\tau$, and thus the estimates of $k\phi$, to be artificially low.

Figs. \ref{fig:flow rate summary} and \ref{fig:Flow rate Slopes} show the effect of changing $Q_T$. Fig.~\ref{fig:flow rate summary} shows that as flow rate increases, the magnitude of the slope of the transformed pressure trace decreases. Deposition occurs less efficiently at higher flow rates. This is expected, since increasing $Q_T$ affects the magnitude of the slope in Eq.~\ref{eq: model} by increasing the Peclet number.  At higher $\text{Pe}$, the diffusive boundary layer is thinner and less material reaches the wall, resulting in less efficient deposition. However, the change in $\text{Pe}$ alone does not explain the differences between flow rates. After adjusting for $\text{Pe}$ by solving for $k\phi$, we see that the effect of increasing the flow rate is much greater than that of increasing $\text{Pe}$ alone. Fitting using the deposition model shows that $k\phi$ is 5.7 times smaller at $Q_{T}=$ 12 $\mu$L/min compared to 1.2 $\mu$L/min (Fig.~\ref{fig:Flow rate Slopes}). Indeed, the relationship between flow rate and deposition efficiency is described well by a power law with an exponent of $-2/3$.  Interestingly, this result suggests that the absolute rate of deposition ($k\phi Q$) does not depend strongly on flow rate.  $k\phi Q$ increases by less than a factor of 2 as $Q_T$ increases by an order of magnitude, as shown in the inset of Fig.~\ref{fig:Flow rate Slopes} and in Table~\ref{table:flow rates kphiQ}.

\subsubsection{Gel swelling}

Another interesting implication of the deposition model is the estimation of the parameter $k$.  We do not know the volume fraction of crosslinked alginate $\phi$ because not all alginate crosslinks within the channel; microscopy shows that some alginate crosslinks after eluting from the channel.  Nevertheless, we can estimate the volume fraction of uncrosslinked alginate by $C_{Alg}/\rho_{Alg}$.  If we assume that $\phi$ is at its maximum, that is that the volume fraction of crosslinked alginate matches that of the uncrosslinked alginate, $\phi = C_{Alg}/\rho_{Alg}$,  we can place a lower limit on parameter  $k$. In Fig. 3, the concentration of alginate, and thus $C_{Alg}/\rho_{Alg}$, is held constant when either calcium (Fig. 3b) or flow rate (Fig. 3f) is varied.  In these cases, estimates of $k$ are directly proportional to the behavior seen in the plots of $k\phi$.  In Fig.~\ref{fig:Concentrations Slopes}, however, the volume fraction of alginate varies. The inset in Fig.~\ref{fig:Concentrations Slopes} shows $k\phi/(C_{Alg}/\rho_{Alg})$ as a function of gel concentration.  Interestingly, the resulting estimates of $k$ values are nearly or greater than 1. Recall that $k$ reflects deposition efficiency based on a volumetric argument. For a solid substance, we would expect a $k$ value less than 1.  An effective $k>1$ suggests that the alginate volume expands during deposition. This is the expected result for a hydrogel, since water is trapped in the polymer matrix.

We can take this argument one step further and estimate the swelling of the gel deposit compared to the volume of alginate passing through the channel.  The volume of the gel obstructing the channel is $\pi (R_0^2-R^2)L = V(1-\bar{R}^2)$. The non-dimensionalized radius $\bar{R}^2$ can be expressed as a function of $\tau$ using Eq. 3: $1-\bar R^2 = B \text{Pe}^{-2/3}k\phi \tau$. The volume of alginate which passes through the channel during a single deposition event is $Q(t-t_{0i})C_{Alg}/\rho_{Alg} = V\tau C_{Alg}/\rho_{Alg}$.  By taking the ratio of these two values, we arrive at a definition of $S^*$:

\begin{equation}\label{eq: Swelling Ratio}
S^* = 
\frac{V (1-\bar R^2)}{ V\tau C_{Alg}/\rho_{Alg}} 
= \frac{B \text{Pe}^{-2/3}k\phi}{ C_{Alg}/\rho_{Alg}}
\end{equation}

We denote this value as $S^*$ to acknowledge that this is similar to a swelling ratio $S$, which is typically defined in bulk gels at rest.  That is, $S$ is the mass (or volume) of a hydrated gel divided by the mass (or volume) of the un-hydrated polymer.  However, Eq. 5 likely underestimates the swelling of the gel deposited in flow.  That is, the denominator of $S^*$ considers the volume of all alginate which flows through the pore.  However, only a fraction of the alginate cross-links and is available for deposition, as discussed previously.  Average $S^*$ values for all experimental conditions can be found in Table \ref{table:flow rates kphiQ}. $S^*$ is greater than 1 at every condition tested and is nearly 100 in some cases. That is, the volume of the deposit is always larger, and often much larger, than expected based on the volume of polymer alone.

The behavior seen in Table \ref{table:flow rates kphiQ} shows the interaction between component concentrations and flow conditions on the final structure and properties of the hydrogel. 
Measurements in the bulk suggest that increasing component concentrations provides internal structure to any gel, thus increasing its stiffness, and also reducing the swelling ratio \cite{davidovich2010quantitative}. 
Interestingly, increasing calcium concentration $C_{Ca^{2+}}$ increases $S^*$ significantly.  This difference may be due to an increase in the deposition rate, that is, the true $k$ value may be higher at elevated $C_{Ca^{2+}}$.  Alternatively, the stiffer gels may be more able to withstand the collapsing effect of shear stress, thereby taking up more space.
However, changing the concentration of gel $C_{Gel}$ does not change $S^*$ nearly as much, showing that the crosslinking density is a more important factor than overall concentration.  
At identical component concentrations but changing flow rates, we find that gels formed in higher flow rates have a lower $S^*$. The relationship fits a power law well, where $S^* \sim Q_T^{-4/3}$. This matches the expected behavior for $S^*$. Equation \ref{eq: Swelling Ratio} shows both $S^*\sim\text{Pe}^{-2/3}$, where $\text{Pe}\sim Q_T$, and also $S^*\sim\text k\phi$.  Independently,  $k\phi \sim Q_T^{-2/3}$ according to Fig.~\ref{fig:Flow rate Slopes}.  As such, the diffusively-driven deposition model fully captures the variation seen in $S^*$ with respect to changing flow rates. 


\subsubsection{Relationship between gel deposition and time to ablation}

The deposition model makes no assumptions about the mechanism of ablation. We might anticipate that ablation occurs when a certain degree of occlusion, and a corresponding critical shear stress, is reached within the channel. 
In Eq. 3, the quantity $\left(1-\bar R^2\right)$ refers to the percentage of the cross sectional area occluded at any given time.  When the flow rate is fixed, this degree of occlusion is proportional to product of the variable quantities $k\phi$ and $\tau$.  At the point of ablation, the efficiency of deposition is known by the fit parameter $k\phi$; $\tau_{Abl}$ is directly assessed from the pressure trace measurements.  If a particular degree of occlusion is required to cause ablation, the model predicts:
\begin{equation}\label{eq: tau kphi ideal}
k\phi \sim \left(1-\bar R_{Abl}^2\right)\tau_{Abl}^{-1}
\end{equation}
\noindent This scaling behavior implies $k\phi \sim \tau_{Abl}^n$ with $n=-1$ as long as the degree of occlusion at ablation $\left(1-\bar R_{Abl}^2\right)$ is constant. That is, $k\phi$ and $\tau_{Abl}$ are inversely proportional to each other.  Gels deposited more efficiently ablate sooner than those that deposited less efficiently, if a fixed degree of occlusion is responsible for ablation.

\begin{figure}[hbt!]
    \centering
    \includegraphics[width=\columnwidth]{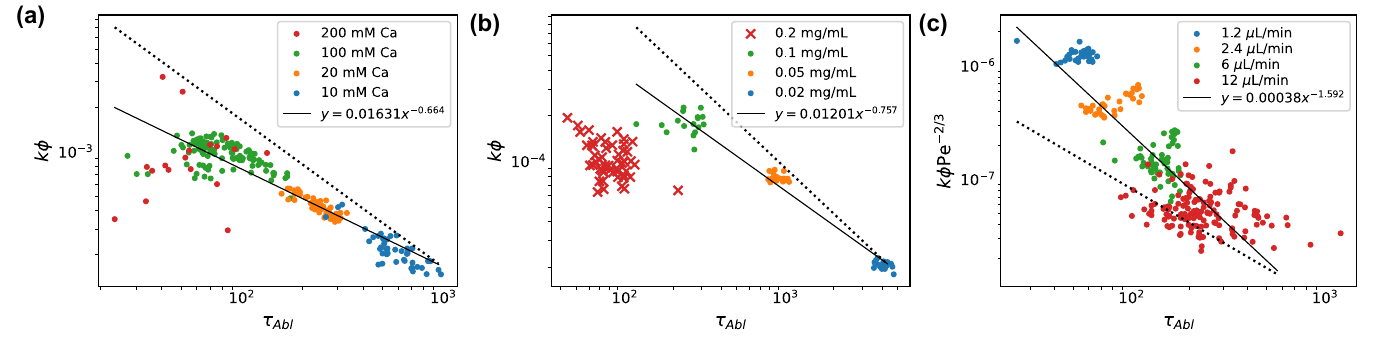}

    \phantomsubfloat{\label{fig:Calcium Collapse}}
    \phantomsubfloat{\label{fig:Concentrations Collapse}}
    \phantomsubfloat{\label{fig:Flow Rates Collapse}}
    \vspace{-2\baselineskip}
        
    \caption[Data Collapse in Model]{
        (\subref{fig:Calcium Collapse}), (\subref{fig:Concentrations Collapse}), (\subref{fig:Flow Rates Collapse}) Deposition rate $k\phi$ as a function of ablation time $\tau_{Abl}$. The solid line represents the power law of best fit. 
        According to the diffusion driven deposition model Eq. \ref{eq: model}, the product between $k\phi$ and $\tau$ is proportional to a specific degree of occlusion $1-\bar R^2$. A representative degree of occlusion is shown in each plot as a dotted line of a power law with an exponent $n=-1$. For experiments where $Q_T$ varies (such as in \subref{fig:Flow Rates Collapse}), we include the factor $ \text{Pe}^{-2/3}$ to maintain the correct proportionality between the axes and the degree of occlusion. }
    \label{fig:Model Collapse}

\end{figure}

Because both $k\phi$ and $\tau_{Abl}$ are known, we can assess the validity of the assumption that a particular degree of occlusion causes ablation. Fig.~\ref{fig:Model Collapse} shows $k\phi$ plotted with respect to $\tau_{Abl}$ on a log-log scale for each data series. Fig.~\ref{fig:Calcium Collapse}, shows various concentrations of calcium at fixed flow rate $Q_T$.  When $C_{Ca^{2+}}$ is varied, the data combined across conditions fit a power law $k\phi \sim \tau_{Abl}^n$ with exponent $n=-0.69$.  Fig. \ref{fig:Concentrations Collapse} shows different values of $C_{Gel}$.  The higher frequency pressure trace features of the highest gel concentration, $C_{Gel} = 0.2$ mg/mL shown in red, exceed the capability of the flow controller as described above. Therefore, these data points are excluded from the power law fit. The exponent of the power law fit in Fig. \ref{fig:Concentrations Collapse} is $n=-0.76$, which is similar to the calcium case.

The dotted lines plotted in Fig.~\ref{fig:Model Collapse} represents ablation that occurs at a fixed degree of occlusion, in which $k\phi \sim \tau_{Abl}^{-1}$.  In Fig.~\ref{fig:Calcium Collapse}, the dotted line corresponds to a degree of occlusion $\left(1-\bar R_{Abl}^2\right) \sim 80$ \%.  Data points falling below the dotted line represent events with less occlusion at the time of ablation than those at the line.  Data collected at $C_{Ca^{2+}}=10$ mM appear closest to the line of $k\phi \sim \tau_{Abl}^{-1}$.  As the amount of calcium increases, the data falls further below the dotted line.  That is, a greater degree of calcium leads to a lesser degree of occlusion at the time of ablation.  When $C_{Ca^{2+}}=100$ mM, in green, the degree of occlusion drops to 44 \%. Remarkably, with the exception of $C_{Ca^{2+}}=200$ mM, the power law behavior with exponent $n=-0.69$ holds within each set of experiments in addition to describing the combined data sets together.  This suggests that, even for gels that deposit in similar chemical conditions, those that deposit more efficiently (larger $k\phi$) occlude the channel less at the moment of ablation than those that deposit less efficiently.  Said in another way, clogs which grow more slowly or with less efficient deposition occlude more before ablating.  

This result holds in Fig.~\ref{fig:Concentrations Collapse} as well, where $n=-0.69$ when considering all data sets combined.  Here the results suggest that gels with higher alginate content occlude the channel less at the moment of ablation than those formed with less alginate.  The dotted line represents a constant degree of occlusion $\left(1-\bar R_{Abl}^2\right) \sim 44$ \%.  Gels formed with $C_{Gel}=0.02$ mg/mL occlude the channel approximately this much at the moment of ablation.  Gels formed with more $C_{Gel}=0.1$ mg/mL, occlude the channel to only 22\%. However, in these data sets, there is less spread of data points within a particular value of $C_{Gel}$.  

The combined results seen in Fig.~\ref{fig:Calcium Collapse} and Fig.~\ref{fig:Concentrations Collapse}, show that, as chemistry changes, $k\phi\sim\tau_{Abl}^{n}$, where $n\sim-0.75$. Whenever either calcium content or alginate content is increased, the degree of occlusion at the moment of ablation decreases.  This conclusion also holds in different channel geometries.  As seen in the  Supplemental Material Fig.~S3, channel size does not alter the scaling behavior seen in Fig.~\ref{fig:Model Collapse}. Further, varying calcium suggests that gels which deposit less efficiently, or, perhaps, more slowly, may occlude more the channel before ablating. The degree of occlusion at ablation depends on deposition efficiency even for gels formed in the same chemical and physical conditions.

Varying $Q_T$ requires a slightly different set of axes.  Fig. \ref{fig:Flow Rates Collapse} shows the expected behavior for constant occlusion at ablation when varying flow rate. The deposition model in Eq. \ref{eq: model} suggests that the flow-rate dependent factor $\text{Pe}^{-2/3}$ modifies the proportionality between $k\phi$ and $\tau$. Therefore, we include it in Fig. \ref{fig:Flow Rates Collapse} so that the dotted line of $n=-1$ is still proportional to a specific degree of occlusion $(1-\bar R^2)$. In Fig. \ref{fig:Flow Rates Collapse}, data points taken the slowest flow rate $Q_T=1.2\mu$L/min occupies the upper left area while those at the largest flow rate $Q_T=12.0\mu$L/min fall in the lower right. The flow rate data fits a power law of $n=-1.59$. That is, when we increase the flow rate, clogs ablate at a lower degree of occlusion. At the fastest flow rate, $Q_T=12 \mu$L/min, in red, the average degree of occlusion is $\sim4$\%. Decreasing the flow rate allows the gel to occlude the channel to a greater degree before ablation occurs.  At the slowest flow rate, $Q_T=1.2 \mu$L/min, in blue, the average degree of occlusion increases to nearly $\sim$25\%.

\subsection{Shear Stress Analysis}

The suggestion that gel properties provide some explanation for the deposition and ablation dynamics leads us to consider the critical shear stress at the point of ablation. In addition to the deposition model, the pressure trace data gathered from the flow tests allows the estimation of shear stress during the gel deposition, particularly at the point of gel ablation where shear stress is responsible for removing the gel from the wall. The increase in pressure during gel deposition is caused by the constriction of the cross-sectional area of the channel.  The constriction of the channel in turn leads to higher flow velocities to maintain the constant flow rate $Q_T$.  As such, measurements of $\Delta P$ can be used to describe the changes in the linear velocity of the flow and viscous shear stresses experienced by the gel. 
As seen in Eq. \ref{eq: Poiseuille}, at constant $Q_T$, an increase in $\Delta P$ reflects a reduction in $R$.  The shear rate at the wall of fluid flowing in a pipe, $\dot{\gamma}$, is described by 
\begin{equation}\label{eq: shear rate in a pipe}
    \dot{\gamma}(t) = \frac{4Q}{\pi R^3(t)}
\end{equation}
Because the alginate solutions are very dilute, with a viscosity near that of water, and because deposition only depletes the flowing fluid of polymer, we use the Newtonian definition for shear stress: $\sigma=\mu\dot{\gamma}$. By following these relations, we arrive at the following relationship between shear stress  and driving pressure $\Delta P$:

\begin{equation}
    \label{eq: shear relationship}
    \frac{\sigma(t)}{\sigma_0} =\frac{\dot{\gamma}(t)}{\dot{\gamma}_0}=\frac{R_0^3}{R(t)^3} = \left(\frac{\Delta P(t)}{\Delta P_0}\right)^{3/4}
\end{equation}

\noindent where the subscript 0 indicates the baseline value in a clean channel. For a clean pipe with an equivalent radius of 19 $\mu$m, the estimated baseline shear stress at a flow rate of 1.2 $\mu$L/min is $\sigma_0=3.7$ Pa. Using the pressure trace and baseline shear stress, we can calculate the shear stress experienced by the clog at any time up to and including the point of ablation:
\begin{equation}\label{eq: shearstress}
    \sigma(t)=\sigma_0 \left(\frac{\Delta P(t)}{\Delta P_0}\right)^{3/4} 
\end{equation}

\begin{figure}
    
    \includegraphics[width=\columnwidth]{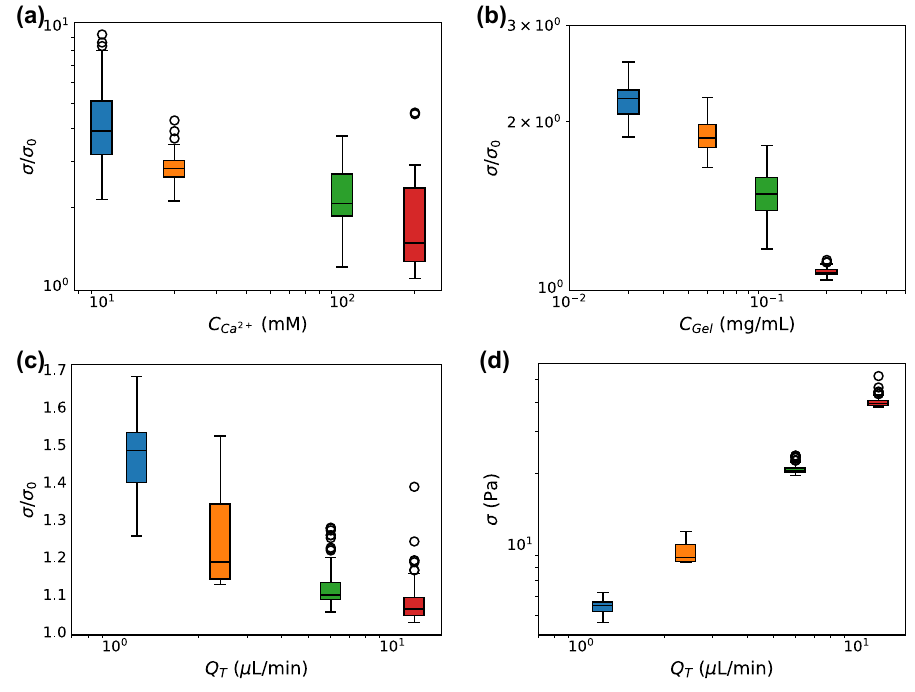} 

    \phantomsubfloat{\label{fig:Calcium Normalized Shear Stress}}
    \phantomsubfloat{\label{fig:Concentrations normalized shear stress}}
    \phantomsubfloat{\label{fig:Flow rate normalized shear stress}}
    \phantomsubfloat{\label{fig:Flow rate shear stress}}
    \vspace{-2\baselineskip}
    
    \caption[Shear Stress Analysis]{
    Shear stress at gel ablation at: 
    (\subref{fig:Calcium Normalized Shear Stress}) different calcium concentrations,
    (\subref{fig:Concentrations normalized shear stress}) different alginate concentrations,
    (\subref{fig:Flow rate normalized shear stress}) various flow rates (normalized stress),
    (\subref{fig:Flow rate shear stress}) various flow rates (absolute stress).
    Baseline shear stress ($\sigma_0$) is calculated to be 3.7Pa, 7.4Pa, 18.6Pa, and 37.1Pa for 1.2, 2.4, 6.0, and 12.0 $\mu$L/min respectively.
    Note: (\subref{fig:Flow rate shear stress}) is plotted on a log-log scale to illustrate the spread of the data more clearly.
    }
    \label{fig:Shear Stress Analysis}
\end{figure}

Fig.~\ref{fig:Shear Stress Analysis} shows the shear stress at ablation for all experimental conditions using  Eq. \ref{eq: shear relationship}.   Fig.s~\ref{fig:Calcium Normalized Shear Stress}, \ref{fig:Concentrations normalized shear stress}, and \ref{fig:Flow rate normalized shear stress} show the effects of $C_{Ca^{2+}}$,$C_{Gel}$, and $Q_T$ respectively on the normalized shear stress, $\sigma/\sigma_0$.  Because the baseline shear stress $\sigma_0$ changes with $Q_T$, we also show the effect of $Q_T$ on the absolute shear stress in Fig.~\ref{fig:Flow rate shear stress} using Eq. \ref{eq: shearstress}. 

As seen in Fig.s ~\ref{fig:Calcium Normalized Shear Stress} and \ref{fig:Concentrations normalized shear stress}, $C_{Ca^{2+}}$ and $C_{Gel}$ affect the shear stress at ablation in a similar way. At lower concentrations of either cross-linker or gel, more pressure and thus more shear stress is required to remove the deposit from the wall of the channel.  As $C_{Ca2+}$ increases by a factor of 20, the mean $\sigma/\sigma_0$ decreases from $4.47$ to $2.1$.  The decrease in $\sigma/\sigma_0$ with $C_{Gel}$ is a similar scale, decreasing from $2.21$ to $1.07$ as $C_{Gel}$ increases by an order of magnitude.  At high $C_{Gel}$ in particular, the gel is able to withstand very little stress before ablation: $\sigma$ is only $\sim 7$ \% larger than $\sigma_0$.

At first glance, a similar trend appears when varying flow rates, with gel deposits formed in faster flow rates withstanding less shear stress than those formed in slower flow rates, as seen in Fig~\ref{fig:Flow rate normalized shear stress}. The normalized shear stress at ablation decreases slightly, from an average of $1.44 \pm 0.016$ to $1.07 \pm 0.003$ for the cases of 1.2 $\mu$L/min and 12 $\mu$L/min respectively. However, since the baseline shear stress is proportionally higher at higher flow rates, the absolute shear stress experienced by the gel at ablation is an order of magnitude higher for gel deposits that form at $Q=12 \mu$ L / min versus 1.2 $\mu$ L / min, as seen in Fig.~\ref{fig:Flow rate shear stress}. $\sigma = 39.6\pm0.13$ Pa at $Q=12 \mu$ L / min, while $\sigma = 5.35\pm0.06$ Pa at $Q=1.2 \mu$ L / min.  Thus, faster flow rates produce gel deposits which are more resistant to ablation.


\section{Bulk Rheology}

\begin{figure}[hbt!]
    \centering
    \includegraphics[width=\columnwidth]{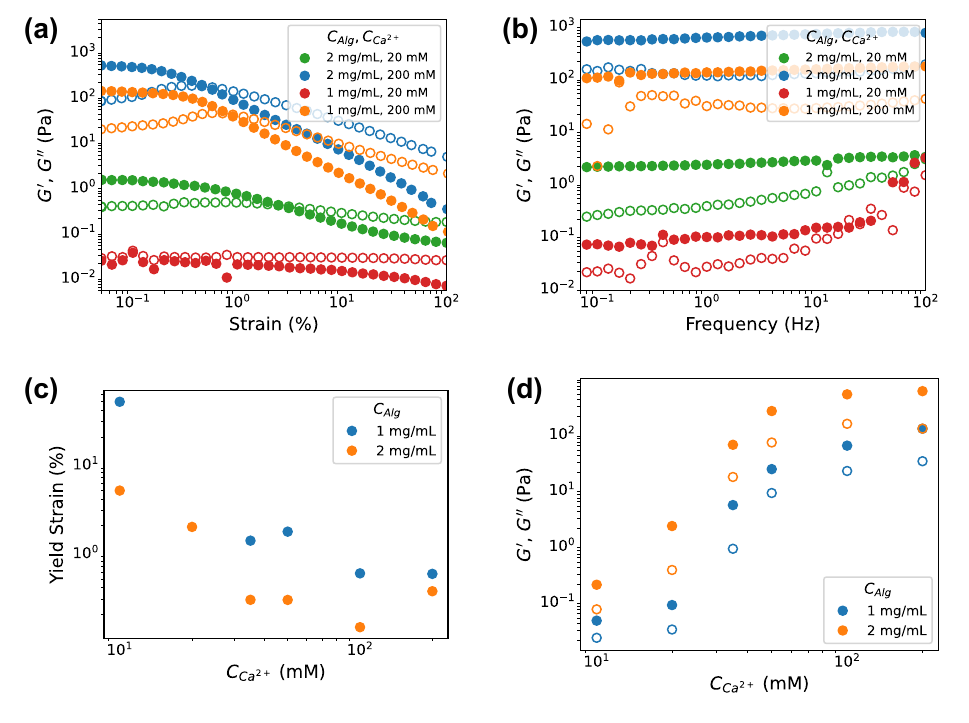} 

    \phantomsubfloat{\label{fig:Rheo amp sweep}}
    \phantomsubfloat{\label{fig:Rheo freq sweep}}
    \phantomsubfloat{\label{fig:Rheo yield strain}}
    \phantomsubfloat{\label{fig:Rheo freq plateau}}
    \vspace{-2\baselineskip}
    
    \caption[Rheology]{ Oscillatory rheology: in all plots, open symbols represent the loss modulus and closed symbols the storage modulus. Representative small amplitude oscillatory shear tests are performed at constant frequency (in (a)) and at constant strain amplitude (in (b)).  Panel (c) shows the yield strain, the strain at which the elastic modulus decreases below the loss modulus. Panel (d) shows the average moduli in the plateau (between 0.1-10 Hz) of the frequency sweep data for alginate concentrations of 1 and 2mg/mL and calcium concentrations ranging  10-200mM.
    }
    \label{fig: Rheology}
\end{figure}

The above analyses suggest that gels which form at higher component concentrations deposit more efficiently, ablate more frequently, and ablate at smaller applied shear stresses.  The shear stress at ablation might indicate either the strength of the adhesion of the gel to the channel or the yield stress of the gel itself.  That is, at the moment of ablation, the flow either pulls the gel off of the wall, or pulls the gel apart from itself.  We can test this intuition by examining the rheological properties of the gels.  
However, the solutions eluted from the fluidic device contain gel rods at dilute concentrations \cite{smith2024}.  Given the dilute nature of the eluted solutions, their viscosity is similar to the injected solutions and exhibit Newtonian behavior only.  Therefore, to quantify the viscoelastic behavior of the gels formed within the microfluidic devices, we increase the overall concentration of the alginate and perform bulk rheology measurements.  We investigate calcium crosslinking ratios over a range that overlaps with ratios experienced in the flow tests and extends to even lower concentrations of calcium.  Figure \ref{fig: Rheology} shows the results of these bulk rheology tests for samples with $C_{Alg}=1$ mg/mL in  and $C_{Alg}=2$ mg/mL: (a) shows representative constant frequency measurements as a function of strain, and (b) shows representative frequency sweep measurements obtained at 0.1\% strain. Figure \ref{fig: Rheology} (c) shows the average moduli in the plateau of the frequency sweep data which occurs between 0.1-10Hz for both alginate concentrations and calcium concentrations of 10-200mM. 

In Figure \ref{fig: Rheology} (a), the strain sweeps show that all samples exhibit a linear regime at very low strain.  All samples show linearity below $\gamma\sim0.1$\%. 
To capture the linear behavior in all samples tested, strain is held constant at 0.1\% to perform frequency sweep tests. Figure \ref{fig: Rheology} (b) shows representative traces of these tests. In the frequency sweeps, all samples exhibit a linear region below a frequency of $\sim10$Hz where the  storage modulus is higher than the loss modulus, indicating gels with more solid-like than liquid-like response. 

In Figure \ref{fig:Rheo yield strain}, we investigate the yield strain, $\gamma_y$, where the liquid-like response overcomes the solid-like response in the strain amplitude sweeps shown in Figure \ref{fig:Rheo amp sweep}.  At the lower concentration of alginate, $C_{Alg}=1$ mg/mL, in blue, and calcium concentrations less than 30 mM, some of the solutions are too dilute to provide a reasonable estimate of $\gamma_y$. For instance, the solution with $C_{Alg}=1$ mg/mL and $C_{Ca^{2+}}=20$ mM, shown in red in Figure \ref{fig:Rheo amp sweep}, does not have a representative data point shown in Figure \ref{fig:Rheo yield strain}. 
At $C_{Alg}=1$ mg/mL, when $C_{Ca^{2+}}<50$ mM, $\gamma_y \sim1$ or 2\%. As the calcium concentration increases to 200 mM, the yield strain decreases to $\gamma_y \sim0.5$\%.  This effect is more pronounced at the higher concentration of alginate.  At $C_{Alg}=2$ mg/mL, $\gamma_y$ decreases by an order of magnitude as calcium content increases from 10 to 200 mM.  Samples with the lowest calcium content exhibit an apparent $\gamma_y$ over an order of magnitude greater than that seen at higher calcium concentrations: $\gamma_y\sim8$\% and 0.3\% at $C_{Ca^{2+}}=10$ and 200 mM, respectively.  In short, as the component concentrations of the gel increase, the yield strain decreases.  While the results in Figure 6 correspond to bulk rheology measurements, we anticipate the same qualitative behavior in flow.  A decrease in yield strain is associated with smaller values of the critical shear stress at the point of ablation, as seen in Figure 5.

Figure \ref{fig: Rheology} (c) shows the average storage and loss moduli in this linear region across all tested samples.
Stiffness increases with more calcium and higher levels of cross-linking. For example, the average storage modulus at $C_{Alg}=2$ mg/mL, $C_{Ca^{2+}}=200$ mM is 570 Pa, while the average storage modulus at $C_{Alg}=2$ mg/mL, $C_{Ca^{2+}}=10$ mM is 0.045 Pa. Similarly, samples with more alginate are stiffer. The average storage modulus at $C_{Alg}=2$ mg/mL, $C_{Ca^{2+}}=200$ mM is 570 Pa, while the average storage modulus at $C_{Alg}=1$ mg/mL, $C_{Ca^{2+}}=200$ mM is 122 Pa. At the lowest calcium concentration (10mM), increasing the overall gel concentration by a factor of 2 increases the storage modulus by nearly an order of magnitude, 0.045Pa and 0.197Pa for  $C_{Alg}=1$ mg/mL and 2mg/mL respectively.

Investigation of the storage modulus suggests that a sufficient amount of calcium can saturate the alginate.  As seen in Figure \ref{fig: Rheology} (d),
both alginate concentrations show an increase in the storage modulus by $>$2 orders of magnitude as calcium concentrations increase from 10mM to 100mM. Increasing calcium concentration from 100mM to 200mM increases the modulus marginally by comparison, 100\% and 14\%  for  $C_{Alg}=1$ mg/mL and 2mg/mL respectively.
These observations may suggest that the ratio of 100 mM calcium per 1 mg/mL alginate may be sufficient to nearly saturate the ability of calcium to crosslink alginate. 

Interestingly, this apparent saturation of the elastic moduli seen in bulk alginate gels might also explain some of the features of the flow tests.  The estimates of deposition efficiency seen in Figure \ref{fig:Calcium Slopes} increase only up to 100 mM of calcium, but not beyond.  The apparent saturation observed in the flow test occurs at a ratio of calcium to alginate beyond 100 mM : 0.1mg/mL.
This ratio in flow is approximately 10 times higher than the crosslinking ratios which lead to sample saturation in the well-mixed bulk gels measured by the rheometer.
Lack of mixing in the device may explain this. Flow tests are run at very high Peclet number, $\text{Pe}>$17000.  As such we should not expect the calcium to be able to diffuse very well out of the calcium stream, except within the diffusive boundary layers near the channel walls or immediately adjacent to the calcium stream. 

\section{Discussion}

Analytical models have been used to describe hydrogel deposition in other contexts.  Two contexts share some similarities with the current study: fibrin mat deposition during blood clot formation and synthetic hydrogel membrane formation.  Blood clot models seek to capture the complexity of the biological system to describe how clot structure and composition vary in 3-dimensional space and time in different physical and biological conditions \cite{Fogelson, chen2019reduced, yesudasan2019recent, nelson2021mathematical}. Like the model presented above, these models must account for both convective and diffusive transport.  However, unlike the model presented above, reaction kinetics during fibrin mat and blood clot formation have a significant impact on deposition behavior \cite{Fogelson, nelson2021mathematical}.  Further, many different protein factors and cell types play a role in the physiology of the coagulation cascade, whose chemical complexity makes it very different from the gelation of a single polymer.

Different models have been proposed to describe \textit{in situ} hydrogel membrane formation \cite{braschler2005gentle,ding2015chip, gargiuli2006microfluidic, johann2007microfluidic, correa2020microfluidic}. In this context, hydrogels are engineered to form a barrier within a microfluidic device, separating two inlet streams and sometimes capturing cells or other biological material within them.  Modeling of this process considers the diffusion of species through the growing membrane to explain the growth rate and final thickness of the gel membrane. However, unlike the current model, the gel membrane is not quantitatively described as significantly occluding the channel and requiring an increase in driving pressure.  Rather, membrane thickness is determined via microscopy rather than externally measurable flow parameters. Additionally, the research in the context of \textit{in situ} flow-assembled membranes typically does not discuss the rheological properties of the gel \cite{ly2021flow}.

The literature on alginate gels, as well as our own rheological results, suggest that increased alginate concentrations and increased calcium cross-linking are associated with increased gel strength \cite{Ouwerx}.
Further, shear forces are less able to disrupt cross-linking and gelation at higher polymer concentrations \cite{Nelson}.
Thus, we expect higher concentrations of alginate and calcium to be associated with stiffer gels in our device. 
Increasing component concentrations leads to faster and more efficient deposition in microfluidic flow, both in our flow tests (see Table \ref{table:flow rates kphiQ}) and with \textit{in situ} membrane formation research  \cite{braschler2005gentle, bazargan2008formation, jia2020microfluidic, correa2020microfluidic}.
While the membrane research shows that increased concentrations lead to larger gel deposits, the gel observed ablation limits the size of gel deposits. Stiffer gels ablate more frequently, likely because the shear stress required for ablation decreases and therefore the gels obstruct flow less before ablation, as seen in Fig.~\ref{fig:Calcium Normalized Shear Stress} and \ref{fig:Concentrations normalized shear stress}. 
Meanwhile, gels formed at lower component concentrations, presumed to be less stiff, are able to grow to occlude more of the channel, and endure greater applied shear stress before ablating from the wall.

In our flow test experiments in which we vary $Q_T$, we find that gels which form in higher shear environments, that is, at higher flow rates, both deposit more efficiently and ablate more quickly.
We also find that gels formed at higher flow rates are much less swollen than those formed at low flow rates, in agreement with similar research \cite{rosella2021microfluidic}. 
Gel growth rate $k\phi Q$ increases slightly with flow rate (Table \ref{table:flow rates kphiQ}). This effect is weak however; increasing the flow rate by an order of magnitude increases $k\phi Q$ by a factor of 2. This is in reasonable agreement with research showing that flow rate only weakly impacts the rate of gel growth \cite{cheng2012biofabrication, luo2010situ, rosella2021microfluidic, ding2015chip, jia2020microfluidic}.  
Our flow tests also show that gels formed at high flow rates can withstand higher shear stress before ablation, as see in Fig.~\ref{fig:Flow rate shear stress}. Theory and previous experimental evidence suggest that bulk gels formed by mixing in high-shear environments are weaker \cite{Omari, Nelson}.
In this way the dependence of ablation on flow rate matches the dependence on concentration: softer gels  withstand higher shear stresses before ablation.
However, the shear stress at ablation of gels formed at higher flow rates is much greater than with gels formed at different chemical conditions. 
Varying component concentrations by an order of magnitude varies the shear stress at ablation by a factor of 2. 
Meanwhile, gels forming at $Q_T$=12.0 $\mu$L/min withstand an order of magnitude more shear stress than those forming at $Q_T$=1.2 $\mu$L/min. 
Because increasing $Q_T$ by an order of magnitude also proportionally increases the applied shear stress, the degree of occlusion actually decreases. The normalized shear stress at ablation decreases with $Q_T$, as seen in Fig.~\ref{fig:Flow rate normalized shear stress}.
So, while gels formed in higher flow rates experience higher shear stresses before ablation, they occlude less of the channel.
In the case of component concentrations, the weaker gels which withstand higher shear stresses occlude the channel more. 

We find that gels which grow more slowly obstruct more before ablation, that is, they ablate at higher shear stresses.  Remarkably, the softest gels can occlude the channel cross sectional area as much as 80\%.  Fig. \ref{fig:Rheo yield strain} shows that softer alginate gels have a higher yield strain than stiffer gels.  So, we find that gels with higher yield strains are able to obstruct the channel more before they ablate.  This relationship suggests a mechanism for the difference in ablation shear stresses. Soft gels are better able to change configuration to minimize the impact of shear stress, and thereby prevent the flow from breaking the gel apart.
Interestingly, the shear stresses observed in this work overlap with certain regions of the cardiovascular system, including arterioles, cerebral vessels, and atherosclerotic arteries \cite{Dolan, sakariassen2015impact}.

\section{Conclusion}

We flow alginate and calcium solutions at constant flow rates through a microfluidic device in which the inlet streams meet at a Y-shaped junction.  The meeting of the two inlet streams results in instantaneous crosslinking: a gel forms downstream of the junction and deposits on the channel walls. The gelled deposit grows to occlude the channel until shear stress pulls it off the wall, ablating the gel. Gel growth and ablation occur regularly and persistently. Varying the flow rate and reaction conditions controls both the magnitude of gel occlusion and frequency of gel ablation. Gels formed at higher component concentrations and slower flow rates exhibit larger increases in pressure and ablate less frequently. These parameters also control the stiffness of the gel formed.  Using analyses based on Poiseuille flow and on diffusively driven deposition, we estimate both the critical shear stress at ablation and the deposition efficiency of the gel reaching and sticking to the wall.  Gels formed at higher component concentrations deposit more efficiently and ablate at lower shear stresses. Complementary rheology shows that these gels also have higher elastic moduli. In all, our results suggest that chemical and physical conditions associated with softer gels result in gels which both deposit more slowly and withstand higher shear stresses before ablation.  This newfound understanding that flow test conditions can simultaneously control both gel properties and deposition/ablation behavior has implications for using intermittent flows as a method to control the formation of gel rods.

\section*{Supplementary Material}

The supplementary material associated with this paper offers additional detail and context for the analysis presented.

\section*{Acknowledgment}
This work is funded by the NSF CBET 2239742 (CAREER).

\bibliography{references.bib}

\end{document}